\documentclass[prb,aps,amssymb,showpacs,twocolumn]{revtex4}
\usepackage{graphicx}

\begin{document}

\title{A General Theorem Relating the Bulk Topological Number
to Edge States \\
in Two-dimensional Insulators}

\author{Xiao-Liang Qi$^{1,2}$, Yong-Shi Wu$^3$ and Shou-Cheng Zhang$^{2,1}$}

\affiliation{$^1$ Center for Advanced Study, Tsinghua University,
Beijing, 100084, China} \affiliation{$^2$ Department of Physics,
McCullough Building, Stanford University, Stanford, CA 94305-4045}
\affiliation{$^3$ Department of Physics, University of Utah, Salt
Lake City, UT 84112-0830}

\date{\today}

\begin{abstract}
We prove a general theorem on the relation between the bulk
topological quantum number and the edge states in two dimensional
insulators. It is shown that whenever there is a topological order
in bulk, characterized by a non-vanishing Chern number, even if it
is defined for a non-conserved quantity such as spin in the case of
the spin Hall effect, one can always infer the existence of gapless
edge states under certain twisted boundary conditions that allow
tunneling between edges. This relation is robust against disorder
and interactions, and it provides a unified topological
classification of both the quantum (charge) Hall effect and the
quantum spin Hall effect. In addition, it reconciles the apparent
conflict between the stability of bulk topological order and the
instability of gapless edge states in systems with open boundaries
(as known happening in the spin Hall case). The consequences of time
reversal invariance for bulk topological order and edge state
dynamics are further studied in the present framework.
\end{abstract}


\maketitle

\section{Introduction}

Since the discovery of the quantized Hall effect two decades
ago\cite{klitzing1980}, topological quantum number has been well
accepted as an important characterization of quantum many-body
systems. The quantized Hall conductance in the integer quantum
Hall effect is first identified, for non-interacting electrons on
a lattice, as a first Chern number over the magnetic Brillouin
zone by Thouless {\em et al.}\cite{thouless1982}, which is also
known as the TKNN number. However, this characterization does not
apply to systems with impurity and/or interactions, since the
Brillouin zone is not defined for these systems. The study on
momentum space topology\cite{volovik2003} can be viewed as the
developments along this direction. It indeed goes beyond the
non-interacting fermion system, by considering the poles of the
single particle Green function, but it pre-requires the Fermi
liquid property of the system. To define a topological
characterization for general many-body systems, Niu {\em et
al.}\cite{niu1985} introduced the technique of twisted boundary
conditions, and showed that the first Chern number can be
generally defined over the parameter space of twisted phases for
any two dimensional insulator.

One of the important observable consequences of the bulk
topological invariance in quantum Hall systems is the presence of
gapless (chiral) edge states on the boundary of the system. Just
like the existence of gapless Goldstone modes is a characteristic
of the spontaneous breaking of a continuous symmetry, the
existence of gapless edge excitations can be considered as a
manifestation of bulk ``topological order", defined through the
topological quantum numbers. According to Laughlin and Halperin's
gauge argument\cite{laughlin1981,halperin1982}, $n$ branches of
gapless edge states must exist for a system with Hall conductance
$\sigma_H=ne^2/h$, since there must be $n$ electrons transferred
from one edge to the other when a unit flux $2\pi$ is
adiabatically threaded through a cylindrical system. The physical
intuition of this argument inspired later
Hatsugai\cite{Hatsugai1993A,Hatsugai1993B} to demonstrate an
explicit and rigorous relation between the bulk TKNN number and
the dynamics of edge states. He has succeeded in finding a
topological characterization of edge states and then proved that
the edge topological number is equal to the bulk TKNN number.
However, this relation cannot be generalized to more general
many-body systems, where the TKNN number has to be replaced by a
more general Chern number as defined in Ref.\cite{niu1985}.
Therefore, it is important to establish a general theorem, similar
to the ``Goldstone theorem" in the case spontaneous symmetry
breaking, that bridges the bulk topological quantum number
generally defined by the twisted boundary condition, and edge
properties for a generic many-body system. To provide such a
theorem is the main goal of the present paper.

This problem is not merely of academic interest, in view of the
recent proposals of the intrinsic spin Hall
effect\cite{murakami2003,sinova2004,murakami2004B,kane2005a} and
the quantum spin Hall effect\cite{bernevig2005A,qi2005,sheng2005}.
Two distinctly new issues arise in the topological
characterization of the quantized spin Hall effect, compared to
that of the quantized charge Hall effect, calling for a clearer
understanding of the general relation between bulk topology and
edge dynamics. Firstly, spin is in general not conserved due to
spin-orbit coupling and, therefore, the Laughlin-Halperin gauge
argument does not apply straightforwardly. However, by
generalizing the twisted boundary conditions in Ref.\cite{niu1985}
to the spin channel, a Chern number can still be
defined\cite{hatsugai2004,hatsugai2005,sheng2006}. Thus the
physical meaning of such bulk Chern number and its relation to the
edge states in the quantized spin Hall case becomes less clear
compared with the quantum Hall case. Secondly, the spin Hall
systems of present interest respect time-reversal invariance,
which leads to additional constraints on the possible Hamiltonians
and on the edge dynamics. In Ref.\cite{kane2005b}, Kane and Mele
proposed a ``$Z_2$ topological order" for T-invariant systems, in
which the T-invariant systems are classified according to whether
there are an even or odd number of Kramers pairs of edge states on
each edge. Such a $Z_2$ characterization is protected by Kramers
degeneracy and seems to be consistent with recent numerical
results\cite{sheng2005,sheng2006}. However, it remains unclear
whether the $Z_2$ characterization can be generalized to many-body
systems with impurities and/or interactions. Careful study on edge
dynamics\cite{wu2005,xu2005} shows that the edge states can become
gapful with sizable strength of interactions, even for the case of
an odd number of Kramers pairs. This suggests that the $Z_2$
classification may {\em not} be a topological order protected by
the bulk gap alone, but depends on more subtle behavior of the
system.

\begin{figure}[tbp]
\begin{center}
\includegraphics[width=2.5in] {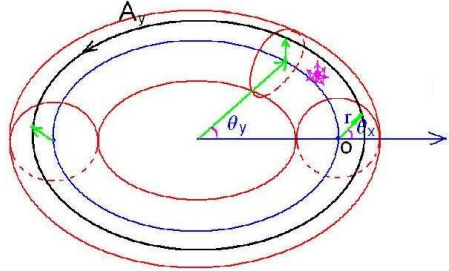}
\end{center}
\caption{Schematic picture of the parameter space torus and the
definition of coordinates $(r,\theta_x,\theta_y)$. A monopole
singularity is drawn in the solid torus. } \label{torus}
\end{figure}

To clarify these issues, in this paper we propose a new framework to
relate the bulk topological number to edge dynamics, by introducing
generalized twisted boundary conditions. In the usual definition of
twisted boundary conditions, the system is defined on a torus
(namely a rectangle with side length L with the opposite boundaries
in $x$- and $y$-directions identified, respectively); a particle
gains an extra phase $e^{i\theta_x}$ (or $e^{i\theta_y}$) whenever
it goes across the boundary in $x$ (or $y$-) direction. (The phases
($\theta_x,\theta_y$) may be associated with the internal degrees of
freedom in the case of the spin Hall effect.) Assuming that the
many-body ground state of the system is separated from the excited
states by a finite gap for all values of the twisted phases
($\theta_x,\theta_y$), the first Chern number is then defined by the
total flux of the Berry-phase gauge field associated with the ground
state over the $(\theta_x,\theta_y)$-space, whose topology is a
torus. To establish a bulk-edge connection, one may relate the
toroidal system to a cylindrical system with open boundaries in the
way as suggested in Ref. \cite{wu1989}. The simplest thought is of
course to cut the torus in real space along a circle, say located at
$x=L$, resulting in a cylinder with two open boundaries at $x=0$ and
$x=L$. Then a particle can no longer go across the boundary at $x=0$
or $x=L$. Conversely, one may view the toroidal system as obtained
by gluing together the two edges of a cylindrical system. In quantum
theory the gluing can be implemented by allowing a tunable
tunnelling between the two edges\cite{wu1989}. This leads to the
idea of changing the twisted phase $e^{i\theta_x}$ of the toroidal
system into a complex number $re^{i\theta_x}$, with $0\leq r \leq
1$. Namely we will allow the hopping amplitude of a particle across
the $x$-boundary to be reduced by a factor $r$. When $r=0$, no
particle can go across the $x$-boundary, and the spatial geometry of
the system becomes a cylinder. With this generalized twisted
boundary condition, the parameter space is now enlarged to three
variables $(r,\theta_x,\theta_y)$. Since at $r=0$, different values
of $\theta_x$ do not make any difference, the geometry of the
parameter space is actually $D^2\times U(1)$, with $(r,\theta_x)$
the polar coordinates of the disk $D^2$. In other words, the
topology of the parameter 3-space is now that of a {\em solid}
torus, the interior of a torus embedded in three dimensions: The
toroidal surface at $r=1$ represents the original toroidal system,
while the central loop at $r=0$ represents the associated
cylindrical system. In this way, the solid torus implements a
continuous interpolation between the toroidal and cylindrical
systems. A schematic picture of the parameter space is shown in Fig.
\ref{torus}. In this framework, the bulk Chern number corresponds to
the total magnetic flux through the toroidal surface at $r=1$, so
inside the solid torus there must be magnetic monopole sources, i.e.
singularities where degeneracy occurs due to the emergence of
gapless edge states and breaks the adiabatic evolution and makes the
Berry phase ill-defined. Consequently, the relation between bulk
topology and edge dynamics is simply dictated by the (magnetic)
Gauss theorem for the solid $(r,\theta_x,\theta_y)$ torus.

Once such a novel connection is established, more complete
understanding on the quantum Hall (QH) and quantum spin Hall (QSH)
systems can be achieved. It will be demonstrated that whenever the
bulk topology is nontrivial, there must be gapless edge states in
some twisted-boundary system with $(r,\theta_x)(r<1)$, but the
gaplessness of edge states does not have to occur in the
open-boundary system with $r=0$. Only when the corresponding
physical quantity in the twisted boundary conditions (charge for QHE
and spin for QSHE) is conserved, can the gapless edge states be sure
to happen in the open-boundary system. The ``shift" of gapless edge
states to generalized twisted boundary conditions provides a natural
rationale to account for the gap opening in the edge channels in the
spin Hall systems\cite{xu2005,wu2005}. Upon considering the time
reversal invariance, some constraints on the winding number
distribution is obtained and the $Z_2$ classification is shown to be
related to the behavior of low energy excitations under time
reversal, not merely to the bulk topology.

The rest of the paper is organized as follows: The proof of the
general theorem on the relation between bulk topology and edge
dynamics is given in Sec. II. Several examples are given in Sec.
III, including the QHE and QSHE cases. The role of time reversal
invariance is discussed and the $Z_2$ classification is analyzed in
Sec. IV. Finally, Sec. V is devoted to summary and discussions.

\section{General Relation between Bulk Topology and Edge States}

\subsection{Generalized twisted boundary conditions and the
bulk Chern number}

There are two equivalent ways to define the twisted boundary
conditions for a toroidal system. One is to impose constraints on
the many body wave function of the form $\Phi({\bf r_1},{\bf
r_2},..)=e^{i\theta}\Phi({\bf r_1} +L{\bf \hat{x}},\bf{r_2},..)$.
The other is to add parameters into the Hamiltonian. Since we would
like to study the Berry phase gauge field under adiabatic evolution,
it is more convenient to choose the latter. For convenience, here we
consider the many-body system on a lattice, with a Hamiltonian of
the form
\begin{eqnarray}
H_0&=&-\sum_{\left\langle
ij\right\rangle}t_{ij,\alpha\beta}c_{i\alpha}^\dagger
c_{j\beta}+H_{\rm int}, \label{originalH}
\end{eqnarray}
in which $c_{i\alpha}$ is the annihilation operator of boson or
fermion at the site $i$, with $\alpha=1,\cdots, s$ labelling
internal degrees of freedom. The first term is the nearest neighbor
hopping term. The second term $H_{\rm int}$ may contain all possible
disorder or interaction terms that conserve the local particle
number: $\left[H_{\rm int},n_i\right]=0, \forall i$. The sites $i$
are defined on a 2-d lattice with size $L$ with periodic boundary
conditions or, in other words, a lattice on a 2-torus $T^2$. Such a
Hamiltonian (\ref{originalH}) describes a wide class of interacting
many-particle systems with nearest neighbor hoppings. Our treatment
below can be generalized to more complex Hamiltonians as discussed
later.

\begin{figure}[tbp]
\begin{center}
\includegraphics[width=2.5in] {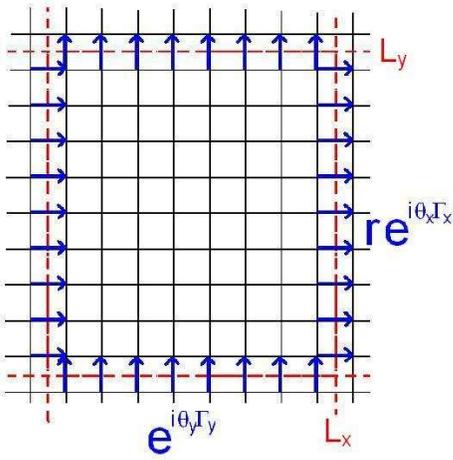}
\end{center}
\caption{Schematic picture of twisted boundary conditions
(\ref{twist}). The $x$ and $y$ boundaries $L_x$ and $L_y$,
respectively, are drawn in red dash lines. Each hopping across them
in the direction indicated by the arrows acquires an additional
matrix factor $re^{i\theta_x\Gamma_x}$ and $e^{i\theta_y\Gamma_y}$,
respectively. The hopping in the opposite direction acquires a
conjugate factor. The factor $r$ is interpreted as amplitude
reduction due to tunnelling between edges.} \label{boundary}
\end{figure}

We introduce on $T^2$ two boundary lines $L_x,L_y$ in the $x$ and
$y$ direction, respectively, as shown in Fig. \ref{boundary}. The
twisted Hamiltonian is defined by changing the hopping matrix
elements across the boundary lines $L_x,L_y$ only in the following
way:
\begin{eqnarray}
t_{ij}\rightarrow \left\{\begin{array}{c c}
t_{ij}r{e^{i\Gamma_{x}\theta_{x}}},&\left\langle
ij\right\rangle\text{ across }L_x\,\\
t_{ij}{e^{i\Gamma_{y}\theta_{y}}},&\left\langle
ij\right\rangle\text{ across }L_y
\end{array}\right\} \,
\label{twist}
\end{eqnarray}
with the parameters $r,\theta_x$ and $\theta_y$ being real. Here the
hopping across boundary lines is referred to that from the left to
the right or from the down to the up, as shown in Fig.
\ref{boundary}. For the hopping across the boundaries in the
opposite direction, we have to take complex conjugate to keep the
Hamiltonian hermitian. All other hopping that do not cross $L_x$ or
$L_y$ remain unchanged. If $r=1$ and $\Gamma_x$ and $\Gamma_y$ are
the unit matrix in internal indices, Eq. (\ref{twist}) describes the
usual twisted boundary conditions introduced in Ref.\cite{niu1985},
with $\theta_x$ and $\theta_y$ the twist phases due to flux
threading.

The above boundary conditions (\ref{twist}) generalize the usual
ones in two important aspects. First we have introduced generators
$\Gamma_x$ and $\Gamma_y$ that act on the internal indices. In the
spin Hall effect, they can be used to describe spin-dependent
twisted phases\cite{sheng2006}. We need to normalize the generators
so that all their eigenvalues are integers so as to make the phase
factor $e^{i\theta_{x(y)}\Gamma_{x(y)}}$ $2\pi$-periodic in
$\theta_x,\theta_y$. The more important feature of our boundary
conditions (\ref{twist}) is that we have introduced a real factor
$r$ that modifies the amplitude of the hopping across the
$x$-boundary. With $r=1$ one recovers the original toroidal system,
while with $r=0$ the particles can not go across or beyond the
$x$-boundaries so that the system becomes a cylindrical one, having
open boundaries. Obviously for the values $0<r<1$, we allow a
tunnelling between the two edges of the cylindrical system. In this
way, our twisted Hamiltonian is parameterized by three real
parameters $r,\theta_x,\theta_y$ and two hermitian generators
$\Gamma_x,\Gamma_y$:
\begin{eqnarray}
H=H_{\Gamma_x\Gamma_y}\left(r,\theta_x,\theta_y\right).
\label{twistedHamiltonian}
\end{eqnarray}
From Eq. (\ref{twist}) it is clear that if $r=0$ all values of
$\theta_x$ are equivalent to each other, so the topology of the
$(r,\theta_x,\theta_y)$-space, with $0 \le r \leq 1$ and $0\le
\theta_x, \theta_y <2\pi$, is a three dimensional {\em solid} torus.
Our twisted Hamiltonian describes a continuous interpolation between
a cylindrical system with $r=0$ and a toroidal system with $r=1$.
The implementation of this interpolation is, as we will see below,
crucial to establishing a general theorem that relates the bulk
topological quantum number of a toroidal system to the edge dynamics
of the associated open-boundary system.

Before proceeding we make the remark that more complex hopping
terms, like next nearest neighbor hopping etc or even two-particle
hopping $c_{i\alpha}^\dagger c_{i\beta}^\dagger c_{j\gamma}
c_{j\delta}\tilde{t}_{\alpha\beta\gamma\delta}^{ij}$, can be
included in the Hamiltonian, as long as all the hoppings are {\em of
a finite range}. The twisted boundary conditions are defined by a
straightforward generalization of Eq. (\ref{twist}) for the hoppings
across the boundaries. All the discussions below will remain
essentially unchanged.

With the parameterized Hamiltonian, the Berry phase gauge field can
be defined at any point $(r,\theta_x,\theta_y)$, where the
Hamiltonian has a unique ground state with an excitation gap. As
usual the Berry gauge potential is given by
\begin{eqnarray}
A_\mu(r,\theta_x,\theta_y)=-i\left\langle
G(r,\theta_x,\theta_y)\right|\frac{\partial}{\partial
\theta_\mu}\left|G(r,\theta_x,\theta_y)\right\rangle ,
\end{eqnarray}
with $\mu=x,y$. The system acquires a Berry phase under an adiabatic
evolution along loop $C$ as the loop integral
\begin{eqnarray}
\Phi(C)=\oint_C {d\theta_\mu A_\mu(r)}.
\end{eqnarray}

For a toroidal system in an insulating phase, we take $r=1$ and
assume that the Hamiltonian
$H_{\Gamma_x\Gamma_y}(1,\theta_x,\theta_y)$ has a non-degenerate
ground state for all $\theta_x,\theta_y$. The bulk first Chern
number is defined as the total flux of the gauge field through the
$(\theta_x,\theta_y)$-torus with $r=1$:
\begin{eqnarray}
N&=&\frac1{2\pi}\int_0^{2\pi}\int_0^{2\pi}
d\theta_xd\theta_yF_{xy}(1,\theta_x,\theta_y)\nonumber\\
&=&\frac1{2\pi}\int_0^{2\pi}\int_0^{2\pi}
d\theta_xd\theta_y\left.\left(\frac{\partial
A_y}{\partial\theta_x}-\frac{\partial
A_x}{\partial\theta_y}\right)\right|_{r=1}. \label{chernnumber}
\end{eqnarray}
Rigorously speaking, the second line is only heuristic, since the
gauge vector potential $A_\mu$ can not be smooth and single-valued
everywhere on the torus (at $r=1$) if $N\neq 0$; otherwise the 2-d
integral in Eq. (\ref{chernnumber}) would vanish due to Stokes'
theorem. A rigorous treatment needs several patches to cover the
torus, together with vector potentials pertaining to each patch.
(For the details, see Ref.\cite{kohmoto1985}.)

\subsection{Proof of the Theorem}

To relate the bulk Chern number to edge states, we start with
re-expressing the Chern number $N$. For any given field strength
$F_{xy}$ on the torus at $r=1$, we choose the gauge in which $A_x$
is single-valued and vanishes. Then we need two patches for $A_y$ to
be single-valued in each. The simplest choice is
\begin{eqnarray}
A^{(1)}_{y}(1,\theta_x,\theta_y)&=&\int_0^{\theta_x}
F_{xy}(1,\theta_x',\theta_y)d\theta_x',
0<\theta_x<2\pi, \nonumber\\
A^{(2)}_{y}(1,\theta_x,\theta_y)
&=&\int_{-\pi}^{\theta_x}F_{xy}(1,\theta_x',\theta_y)
d\theta_x', -\pi< \theta_x<\pi,\nonumber\\
A_x(1,\theta_x,\theta_y)&\equiv & 0 .
\end{eqnarray}
Here $A^{(1)}_y$ and $A^{(2)}_y$ have discontinuity at $\theta_x=0$
and $\theta_x=\pi$, respectively. In this way the torus is covered
by two cylinders defined by $\theta_x\neq 0$ and $\theta_x\neq \pi$.
With this gauge fixing, the Chern number (\ref{chernnumber}) is
expressed as
\begin{eqnarray}
N&=&\frac1{2\pi}\int_{\eta}^{2\pi-\eta}d\theta_x\int_0^{2\pi}
d\theta_y\frac{\partial A^{(1)}_{y}}{\partial \theta_x}\nonumber\\
& &+\frac1{2\pi}\int_{-\eta}^{\eta}d\theta_x\int_0^{2\pi}
d\theta_y\frac{\partial A^{(2)}_{y}}{\partial \theta_x}\nonumber\\
&=&\lim_{\eta\rightarrow 0^+}\frac1{2\pi}\int_{\eta}^{2\pi-\eta}
d\theta_x\frac{\partial}{\partial\theta_x}\left(\int_0^{2\pi}
d\theta_yA^{(1)}_{y}\right). \label{chernnumber2}
\end{eqnarray}
Defining
\begin{eqnarray}
\phi(r,\theta_x)&=&\int_0^{2\pi}d\theta_y
A^{(1)}_{y}(r,\theta_x,\theta_y), \nonumber\\
\Phi(r,\theta_x)&=&\exp\left[i\phi(r,\theta_x)\right],
\label{defineBerryphase}
\end{eqnarray}
then Eq. (\ref{chernnumber2}) can be written as
\begin{eqnarray}
N&=& \frac1{2\pi}\int_{0^+}^{2\pi^-} d\theta_x\frac{\partial
\phi(1,\theta_x)}{\partial\theta_x}
\nonumber \\
&=& -\frac i{2\pi}\oint_{r=1}\Phi^{-1} d\Phi .
\nonumber\\
\label{windingnumber}
\end{eqnarray}
Consequently, there is a $2\pi N$ jump in the $\phi$-field, but the
$U(1)$ phase field $\Phi(r,\theta_x)$ is single-valued everywhere on
the torus at $r=1$. The Chern number is then equivalent to a winding
number $N$ of the $\phi$-field on the loop
$r=1,\theta_x\in[0,2\pi)$. For more general gauge choices, the value
of $\Phi$ can change by a constant phase factor, but the relation
between the Chern number and the winding number remains true as long
as $A_x$ is kept to be single-valued.

The physical meaning of the phase field $\Phi(r,\theta_x)$ is simply
the Berry's phase that the system acquires due to an adiabatic
evolution along the closed path with $\theta_y$ varying from 0 to
$2\pi$ and with $r$ and $\theta_x$ fixed. Since the Hamiltonian
depends on $r$ and $\theta_x$ through $re^{\pm i\Gamma_x\theta_x}$
only, all the parameters $(0,\theta_x)$ corresponds to the same
Hamiltonian, which describes an open-boundary system. Consequently,
the phase factor $\Phi(r,\theta_x)$ is defined on a 2d plane, with
$(r,\theta_x)$ as the polar coordinates.

Now the relation between Chern number and the gapless edge states
becomes clear. If $N\neq 0$, the non-vanishing winding number of
$\phi(r,\theta_x)$ on the unit circle $r=1$ requires vortex-like
singularities in the disk bounded by the unit circle. Recall that
the Berry phase $\Phi(r,\theta_x)$ is always well-defined by Eq.
(\ref{defineBerryphase}) as long as the system
$H_{\Gamma_x\Gamma_y}(r,\theta_x,\theta_y)$ has a unique ground
state and a gapped excitation spectrum for all $\theta_y\in
[0,2\pi)$. Thus we conclude that a singularity of the $\Phi$-field
can only occur at a point $(r,\theta_x)$ where the system becomes
either gapless or has ground state degeneracy for some $\theta_y$.
In summary, we have proved the following theorem:

\begin{itemize}
\item{\underline{Theorem 1}: {Whenever the Chern number defined by
Eq.(\ref{chernnumber}) is non-vanishing, there must be a proper
set of twisted boundary conditions $(r,\theta_x,\theta_y)$ with
$r<1$, for which the Hamiltonian $H_{\Gamma_x\Gamma_y}
(r,\theta_x,\theta_y)$ has either ground state degeneracy or
gapless excitations.}}
\end{itemize}

As mentioned above, the boundary conditions for the system with
$r=1$ correspond to the usual definition of the twisted boundary
conditions with twist phases $\theta_x,\theta_y$, while $r=0$
corresponds to defining a cylindrical system with open boundaries.
The twisted boundary conditions with $0<r<1$ correspond to a (bent)
cylindrical system with a certain amount of inter-edge tunnelling.
The above theorem predicts only the existence of singularities of
the Berry phase gauge field for systems with twisted boundary
condition with $r<1$, due to the appearance of gapless states,
without providing detailed information about where the singularities
are. It is possible that a system with a non-vanishing bulk Chern
number has gapless edge excitations in geometry with open-boundary
(at r=0), like what happens in QH systems. But it is also possible
that the gapless points shift away from $r=0$, which means that the
edge states are gapped in open-boundary systems but become gapless
for some points $(r,\theta_x,\theta_y)$ with $r<1$. The latter case
occurs for some QSH systems, as will be shown in next section. That
both possibilities exist is the major new lesson we learn from
Theorem 1, which improves our understanding of the relations between
bulk topology and edge states. More precisely, we have learned that

\begin{enumerate}
\item {The bulk topological order, defined by first Chern number
(\ref{chernnumber}) and protected by the bulk gap, always manifests
itself by the existence of gapless edge states with a proper amount
of inter-edge tunnelling;} \item {The gaplessness of edge states in
an {\em open-boundary} system is {\em not protected solely} by the
bulk topology; its stability requires additional conditions.}
\end{enumerate}

\section{Examples: Quantum Charge and Spin Hall Effects}

In this section, we will consider the usual integer quantum Hall
effect and the recently proposed quantum spin Hall effect, as two
simplest examples of topological insulators, and study the
consequences of Theorem 1. For the quantum Hall effect, it is shown
that the charge conservation plays an essential role in protecting
gapless edge states in systems with open boundary. For the spin Hall
effect the opposite occurs, i.e. the edge states in an open-boundary
system may be gapped because spin is not conserved due to spin-orbit
coupling. However, as a nontrivial consequence of the bulk
topological order, the edge states can become gapless with
increasing inter-edge tunnelling to a certain ``critical" value. We
will show this explicitly by numerical and analytic methods.

\subsection{Quantum Hall Effect: \\
Role of Charge Conservation}

Compared with more general topological insulators defined by a
nonzero bulk Chern number (\ref{chernnumber}), the distinct feature
of the quantum Hall effect is the charge conservation, i.e. the
twisted boundary conditions are gauge equivalent to flux threading
the torus and the Laughlin-Halperin gauge argument can apply. More
generally, it is shown below that the edge states of an
open-boundary system, $H_{\Gamma_x\Gamma_y}(r=0,\theta_x,\theta_y)$,
are gapless in the thermodynamic limit if the following conservation
conditions are satisfied:
\begin{eqnarray}
\left[\sum_i c_i^\dagger \Gamma_\mu
c_i,H_0\right]&=&0,\qquad(\mu=x,y)\nonumber\\
\left[\Gamma_x,\Gamma_y\right]&=&0,\label{conservation}
\end{eqnarray}
where $H_0$ is the untwisted Hamiltonian (\ref{originalH}) for a
system with periodic boundary conditions.

Under the conditions (\ref{conservation}), one can make a unitary
transformation, $U=\exp\left\{i\frac\theta L\sum_ii_xc_i^\dagger
\Gamma_x c_i\right\}$, to show that the twisted Hamiltonian
(\ref{twistedHamiltonian}) is gauge equivalent to the flux-threading
Hamiltonian,
$\tilde{H}_{\Gamma_x\Gamma_y}(r,\theta_x,\theta_y)\equiv
UH_{\Gamma_x\Gamma_y}(r,\theta_x,\theta_y)U^{\dagger}$. Namely
$\tilde{H}$ can be obtained by replacing
\begin{eqnarray}
t_{ij}\rightarrow t_{ij}e^{i\Gamma_x\frac{\theta_x}{L}(j_x-i_x)},
\qquad \forall \left\langle ij\right\rangle
\end{eqnarray}
in the original Hamiltonian $H_0$. Here $L$ is the size of the
system along $x$ direction. The second equation in Eq.
(\ref{conservation}) is necessary to maintain the conservation of
$\Gamma_x,\Gamma_y$ in the twisted system. The boundary condition in
$y$-direction is kept the same as in eq. (\ref{twist}).
Consequently, the Hamiltonians $H(r,\theta_x,\theta_y)$ and
$\tilde{H}(r,\theta_x,\theta_y)$ have identical energy spectrum. On
the other hand, since $A_x\rightarrow 0$ for $L\rightarrow \infty$,
the energy spectrum of $\tilde{H}$ should be insensitive to
$\theta_x$, except for the special case where the system has phase
stiffness due to off-diagonal-long-range-order. In other words, if
for a value $r_0\neq 0$, the Hamiltonian
$\tilde{H}_{\Gamma_x\Gamma_y}(r_0,\theta_x,\theta_y)$ were gapless
in the thermodynamic limit, then
$\tilde{H}_{\Gamma_x\Gamma_y}(r_0,\theta_x',\theta_y)$ should also
be gapless for all $\theta_x'\in[0,2\pi)$. Thus if a singularity
occurs for $r\neq 0,\theta_x$, then the points on the circle (with
$0\leq \theta_x <2\pi$ and the same $r_0$) would all correspond to
gapless systems, providing a macroscopically large winding number.
Therefore $r_0$ must be zero. So we conclude that the gapless
singularity in parameter space for a system with a nonzero bulk
Chern number $N\neq 0$ in the thermodynamic limit can only occur at
the point $r=0$, corresponding to a system with open boundaries with
gapless edge states.

This argument, perhaps not very rigorous mathematically, is enough
to provide a better understanding of the role of charge conservation
in the bulk-edge relation in the quantum Hall effect. In particular
this implies that the stability of edge states in open-boundary
systems is {\em not} protected by the bulk Chern number alone, if
$\Gamma_x$ or $\Gamma_y$ are not generators of conserved quantities
or if they do not commute with each other. This is exactly what
happens in the case of the quantum spin Hall systems, which we will
study in more details below.

\subsection{Quantum Spin Hall Effect:\\
a Numerical Study}

The bulk topological number in the quantum spin Hall effect
corresponds to choosing the internal generators in the twisted
boundary condition (\ref{twist}) to be
\begin{eqnarray}
\Gamma_x=S,\qquad \Gamma_y=1,
\end{eqnarray}
in the spin channel. Here $S$ is a component of spin or pseudospin
operators. There are two key differences between the QSH and QH
systems; namely (1) $S$ may be {\it non-conserved}; (2) $S$ is odd
under time reversal (TR). In this section, we will focus on the
consequences of point (1), and leave point (2) to next section.

When $S$ is not conserved, it is possible for the singularity in the
$(r,\theta_x)$ plane to shift to some point with $0<r<1$. To see
indeed this can happen in the twisted systems with edge tunnelling,
below we will study the edge dynamics in a generalized version of
the spin Hall system proposed in Ref.\cite{qi2005}. The Hamiltonian
for this system reads
\begin{eqnarray}
H=\sum_{\left\langle ij\right\rangle}c_i^\dagger d_{ij}c_j\equiv
\sum_{\left\langle ij\right\rangle}c_i^\dagger
\left(t+Vd_{ij}^a\Gamma_a \right)c_j,
\end{eqnarray}
in which $\Gamma_a, a=1,2,..,5$ are the generators of $SO(5)$
Clifford algebra, and $d_{ij}^a=\frac{1}{N^2}\sum_{\bf k}d_a({\bf
k})e^{i{\bf k\cdot r_i}}$ with $d_a(k)$ given by
\begin{eqnarray}
d_1(k)&=&-\sqrt{3}c\sin k_y ,\nonumber\\
d_2(k)&=&-\sqrt{3}c\sin k_x ,\nonumber\\
d_3(k)&=&-\sqrt{3}\sin k_x\sin k_y ,\nonumber\\
d_4(k)&=&\sqrt{3}\left(\cos k_x-\cos k_y\right) ,
\nonumber\\
d_5(k)&=&2-e_s-\cos k_x-\cos k_y  . \label{choice1}
\end{eqnarray}
Since we are interested in the spin Hall insulator phase, direct
hopping $t$ is expected to be small. For simplicity, here and below
we will take $t=0$. In the case of inversion symmetry,
$d_1=d_2\equiv 0$ and there is a conserved pseudo-spin operator:
$\Gamma_{12}=\left[\Gamma_1,\Gamma_2\right]/(2i)$. With the
identification $\Gamma_x=\Gamma_{12},\Gamma_y=1$, the twisted
boundary conditions (\ref{twist}) can be expressed as
\begin{eqnarray}
d_{(L,y_i),(1,y_i)}& \rightarrow
&d_{(L,y_i),(1,y_i)}re^{i\theta_x\Gamma_{12}}\,\nonumber\\
d_{(x_i,L),(x_i,1)}&\rightarrow &e^{i\theta_y}d_{(x_i,L),(x_i,1)} \
. \label{twistboundaryex}
\end{eqnarray}
\begin{figure}[tbp]
\begin{center}
\includegraphics[width=2.5in] {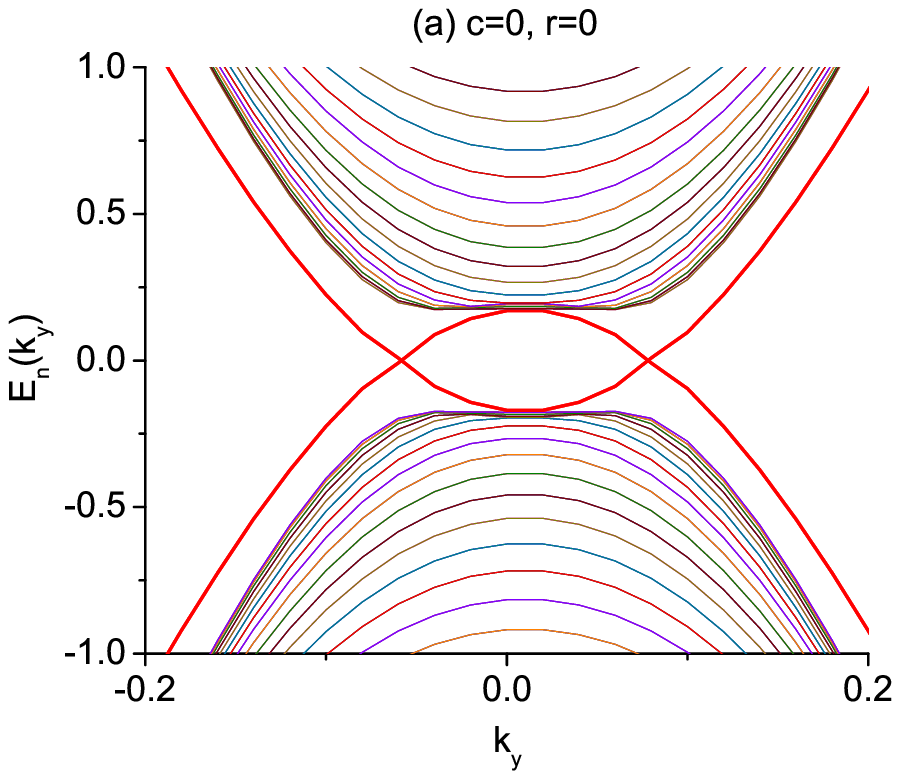}
\end{center}
\begin{center}
\includegraphics[width=2.5in] {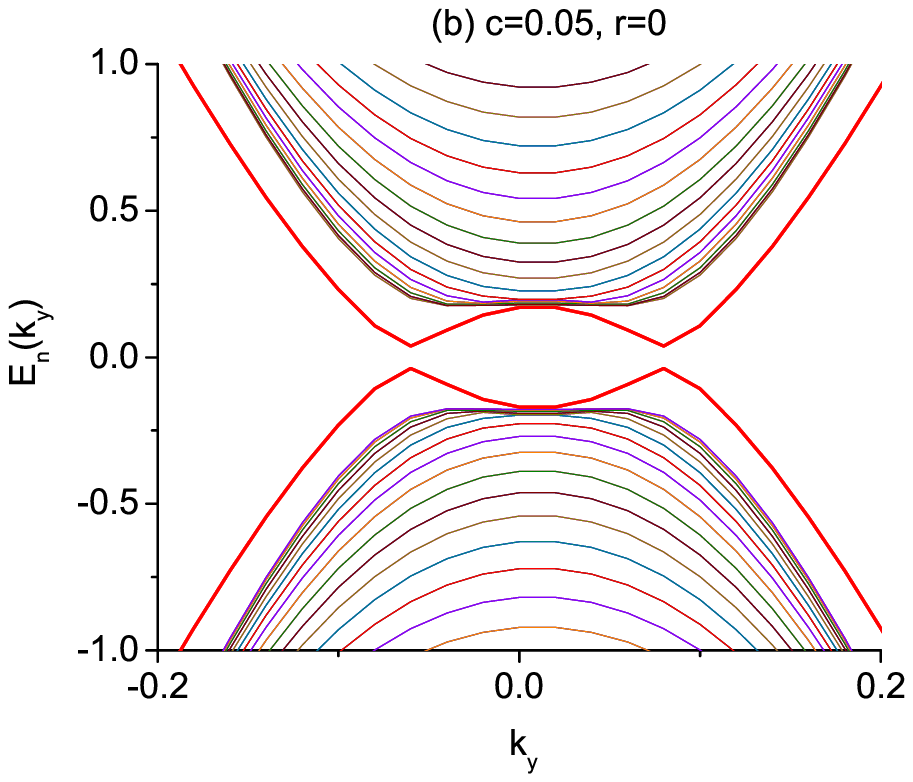}
\end{center}
\begin{center}
\includegraphics[width=2.5in] {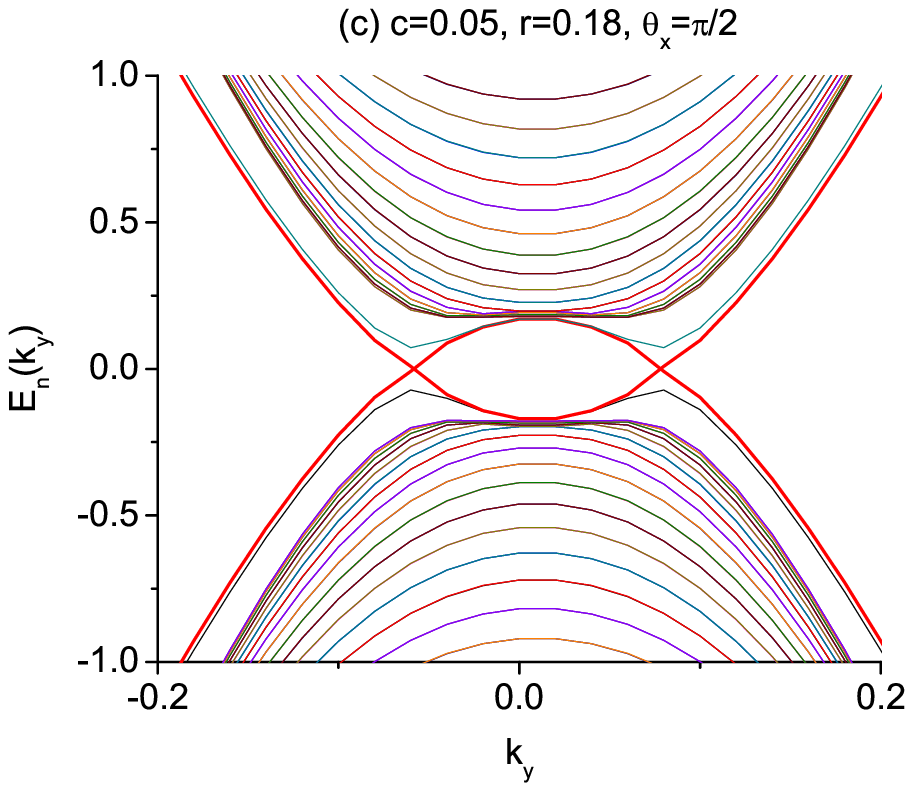}
\end{center}
\caption{The 1-d band structure for (a) $c=0$ and the open-boundary
condition $r=0$; (b) $c=0.05$ and the open-boundary condition $r=0$;
(c) $c=0.05$ and the twisted boundary conditions
$r=0.18,\theta_x=\pm\pi/2,\theta_y=\pi$. }
\label{twistbandstructure}
\end{figure}

Since $\Gamma_{12}$ is conserved, the system is equivalent to a
decoupled bilayer quantum Hall system, and the bulk Chern number
defined for the Hamiltonian
$H_{\Gamma_{12},1}(r=1,\theta_x,\theta_y)$ is $N=4$ for $0<e_s<4$,
as shown in Ref.\cite{qi2005}. According to the arguments in the
last subsection, there should be gapless edge states in the
open-boundary system, as verified numerically in
Fig.\ref{twistbandstructure} (a).

\begin{figure}[tbp]
\begin{center}
\includegraphics[width=2.5in] {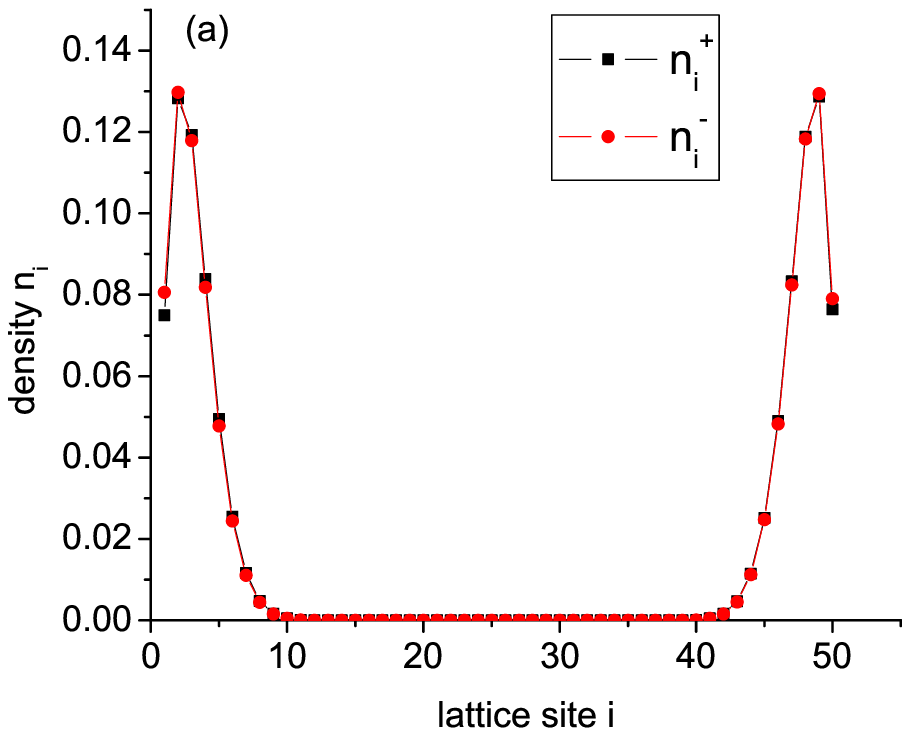}
\end{center}
\begin{center}
\includegraphics[width=2.5in] {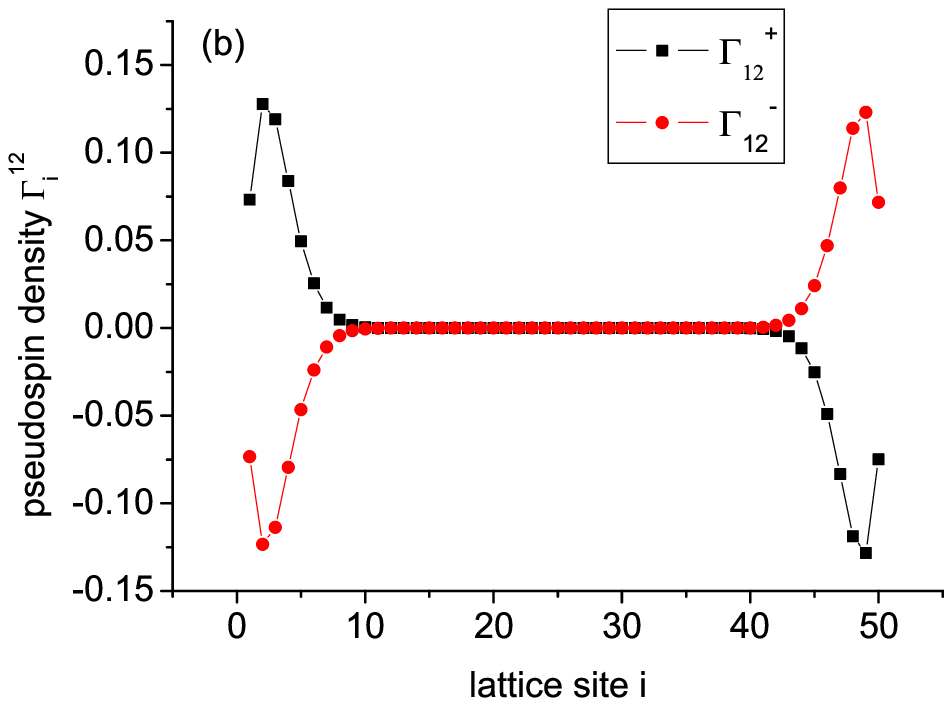}
\end{center}
\caption{The distribution of (a) charge density; (b) pseudo-spin
$\Gamma_{12}$ density of the two edge states with wavevector
$k=0.16\pi$ and with $c=0.05, r=0.18, \theta_x=\pi/2$. }
\label{Dsdist}
\end{figure}

However, when $d_1,d_2\neq 0$, the edge states in the system with
open boundary conditions will become gapful with a gap of the order
$\sqrt{d_1^2+d_2^2}$, since the $\Gamma_1,\Gamma_2$ terms will mix
the states with opposite $\Gamma_{12}$-eigenvalue. Also as long as
$|d_1|,|d_2|\ll \sqrt{d_\alpha d^\alpha},\alpha=3,4,5$, the bulk gap
cannot be closed by turning on $d_1,d_2$. Consequently, the Berry
phase gauge field remains well-defined on the torus with $r=1$, and
thus the Chern number for the Hamiltonian
$H_{\Gamma_{12}1}(r=1,\theta_x,\theta_y)$ remains unchanged.
According to our Theorem 1, {\em there must be some $(r\neq 0,
\theta_x,\theta_y)$ where the system become gapless.} Since all the
twist transformations only make changes in the boundary terms in the
Hamiltonian, it is reasonable to expect the gapless excitations to
live around the edges. The nonzero tunnelling $r\neq 0$ between the
edges actually mixes the original edge states to form new gapless
excitations that are localized at the edge. To verify this, we have
done a numerical calculation for the model with $c=0.05$. We
identified two gapless points in parameter space at
$r=0.18,\theta_x=\pm\pi/2, \theta_y=\pi$. The evolution of the band
structure is shown in Fig. \ref{twistbandstructure}. At each gapless
point there are two level crossings, which keeps the total winding
number $4$. The charge and pseudo-spin density distributions of the
gapless states are shown in Fig. \ref{Dsdist}, confirming that it is
localized symmetrically around the edge, and the pseudo-spin
polarization is opposite on the two sides of the edge.

\subsection{Quantum Spin Hall Effect:\\
a Perturbative Study}

It will be helpful to study how the gapless edge state arises from
the original edge states through tunnelling in perturbation theory.
Under the condition $|d_1|,|d_2|\ll \sqrt{d_\alpha d_\alpha}$ and
$r\ll 1$, all the effects induced by nonzero $d_1,d_2$ and by
twisted boundary conditions are only significant for the edge
states, while leading to little change in the bulk states. Thus it
is reasonable to describe the low energy dynamics in an effective
1-d edge Hamiltonian. Let us first take the continuum limit of the
gapless edge Hamiltonian for the system with $d_1=d_2=0,r=0$ as the
starting point:
\begin{eqnarray}
H^{\rm edge}_0=\sum_{s\alpha}\int dk\alpha
vk\left(\phi^\dagger_{kLs\alpha}\phi_{kLs\alpha}
-\phi^\dagger_{kRs\alpha}\phi_{kRs\alpha}\right),
\end{eqnarray}
in which $\alpha=\pm 1$ corresponds to the eigenvalue of
$\Gamma_{12}$, and $L,R$ the states at the left and right edges.
$s=1,2$ correspond to two crossings in Fig. \ref{twistbandstructure}
(a), and the wave-vector $k$ is defined in reference to the
corresponding crossing points. The time reversal transformation of
the 8 edge states is $T\phi_{k,L(R),1,\alpha}T^{-1}=\alpha
\phi_{-k,L(R),2,-\alpha}$. When $r\neq 0$, we need to incorporate
the mixing between different edges, and when $d_1,d_2\neq 0$ the
mixing between different $\alpha$. The states with different $s$
will never be mixed. So we take the perturbed effective Hamiltonian
to be of the form
\begin{eqnarray}
H_{\rm edge}&=&\int dk \alpha
vk\left(\phi^\dagger_{kLs\alpha}\phi_{kLs\alpha}
-\phi^\dagger_{kRs\alpha}\phi_{kRs\alpha}\right)\nonumber\\
&+&\int dk\left(u_{Ls}\phi_{kLs\uparrow}^\dagger
\phi_{kLs\downarrow}+u_{Rs}\phi_{kRs\uparrow}^\dagger
\phi_{kRs\downarrow}+h.c.\right)\nonumber\\
&+& r\sum_s\int dk\left(\phi_{kLs\alpha}^\dagger
e^{i\theta\alpha}\phi_{kRs\alpha}+h.c.\right)\nonumber\\
&\equiv&\sum_s\int dk\Phi^\dagger_{ksi}M^s_{ij}\Phi_{ksj}, \qquad
(i,j=1,\cdots,4) \label{perturbHamiltonian}
\end{eqnarray}
in which $\Phi_{ks}=\left(\phi_{kLs\uparrow},\phi_{kLs\downarrow},
\phi_{kRs\uparrow},\phi_{kRs\downarrow}\right)$. The matrix elements
$u_{L(R)s}$ are defined as
\[
u_{L(R)s}\equiv \left\langle k=0,L(R)s\uparrow\right|
d_1\Gamma^1+d_2\Gamma^2\left|k=0,L(R)s\downarrow\right\rangle
\]
where the $k$-dependence of the matrix elements is ignored. The
matrix $M^s_{ij}$ is defined as
\begin{eqnarray}
M^s&=&\left(\begin{array}{c c c c}vk & u^*_{Ls} & re^{i\theta} & 0\\
u_{Ls} & -vk & 0 & re^{-i\theta}\\re^{-i\theta} & 0 &-vk &u_{Rs}^*
\\0 & re^{i\theta} & u_{Rs}& vk \end{array}\right) \ .
\end{eqnarray}

Direct diagonalization of $M$ gives the four eigenvalues:
\begin{widetext}
\begin{eqnarray}
E_{s\alpha\beta}=\alpha\sqrt{v^2k^2
+\frac{{|u_{Ls}|}^2+{|u_{Rs}|}^2}2+r^2
+\beta\sqrt{\frac{\left({|u_{Rs}|}^2-{|u_{Ls}|}^2\right)^2}4
+r^2\left({|u_{Ls}|}^2+{|u_{Rs}|}^2
+2{|u_{Ls}|}{|u_{Rs}|}\cos2\tilde{\theta}_s\right)}} \ ,
\end{eqnarray}
\end{widetext}
in which $\alpha,\beta=\pm 1$ and
\[
\tilde{\theta}_s=\theta+{\rm Arg}(u_{Ls})-{\rm Arg}(u_{Rs}) \ .
\]

The equation $E_{s\alpha\beta}=0$ has two solutions:
\begin{eqnarray}
\tilde{\theta}_s=0\text{ or }\pi,\qquad
r=r_s\equiv\sqrt{|u_{Ls}u_{Rs}|} \ .
\end{eqnarray}

\begin{figure}[tbp]
\begin{center}
\includegraphics[width=2.5in] {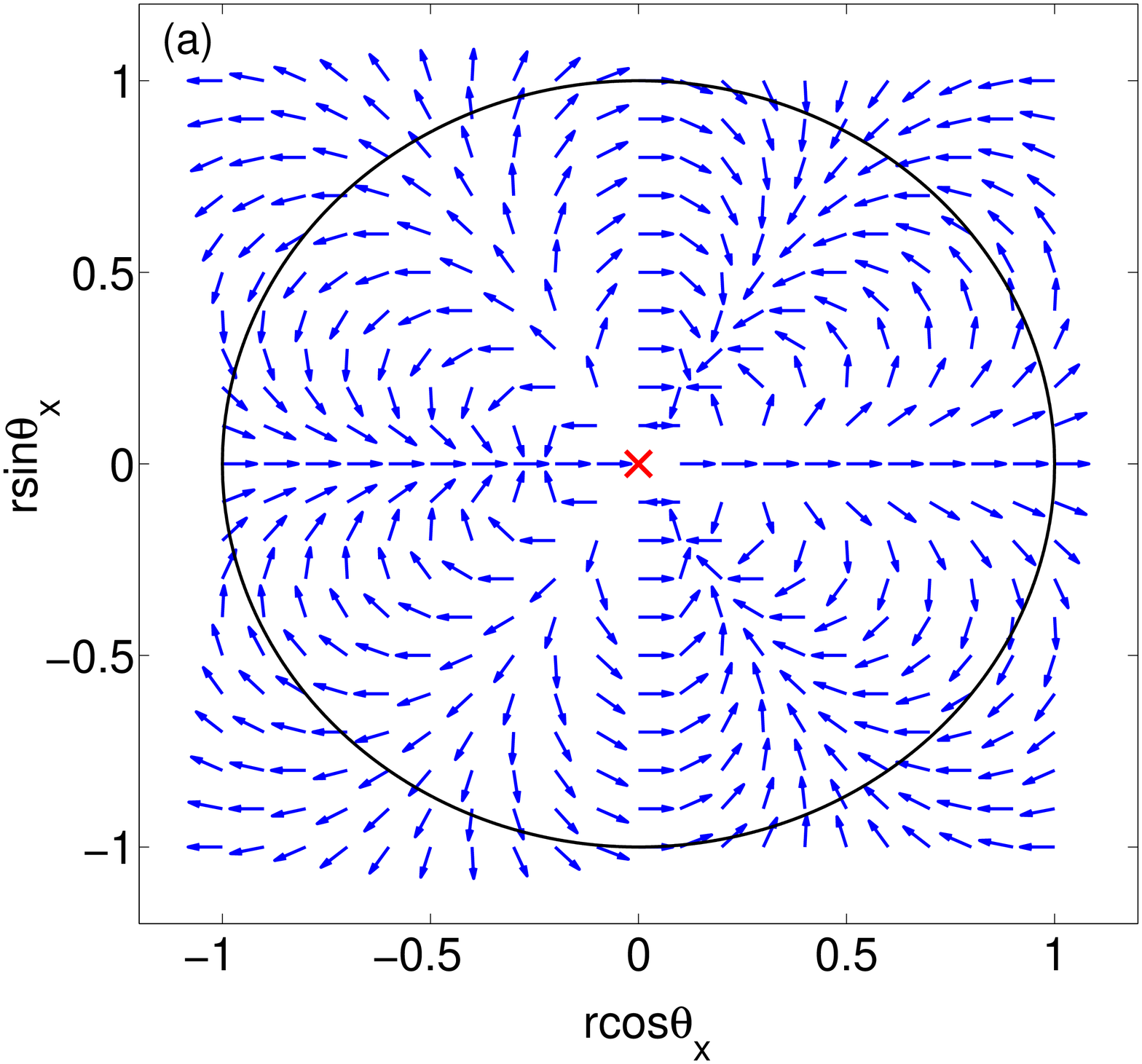}
\end{center}
\begin{center}
\includegraphics[width=2.5in] {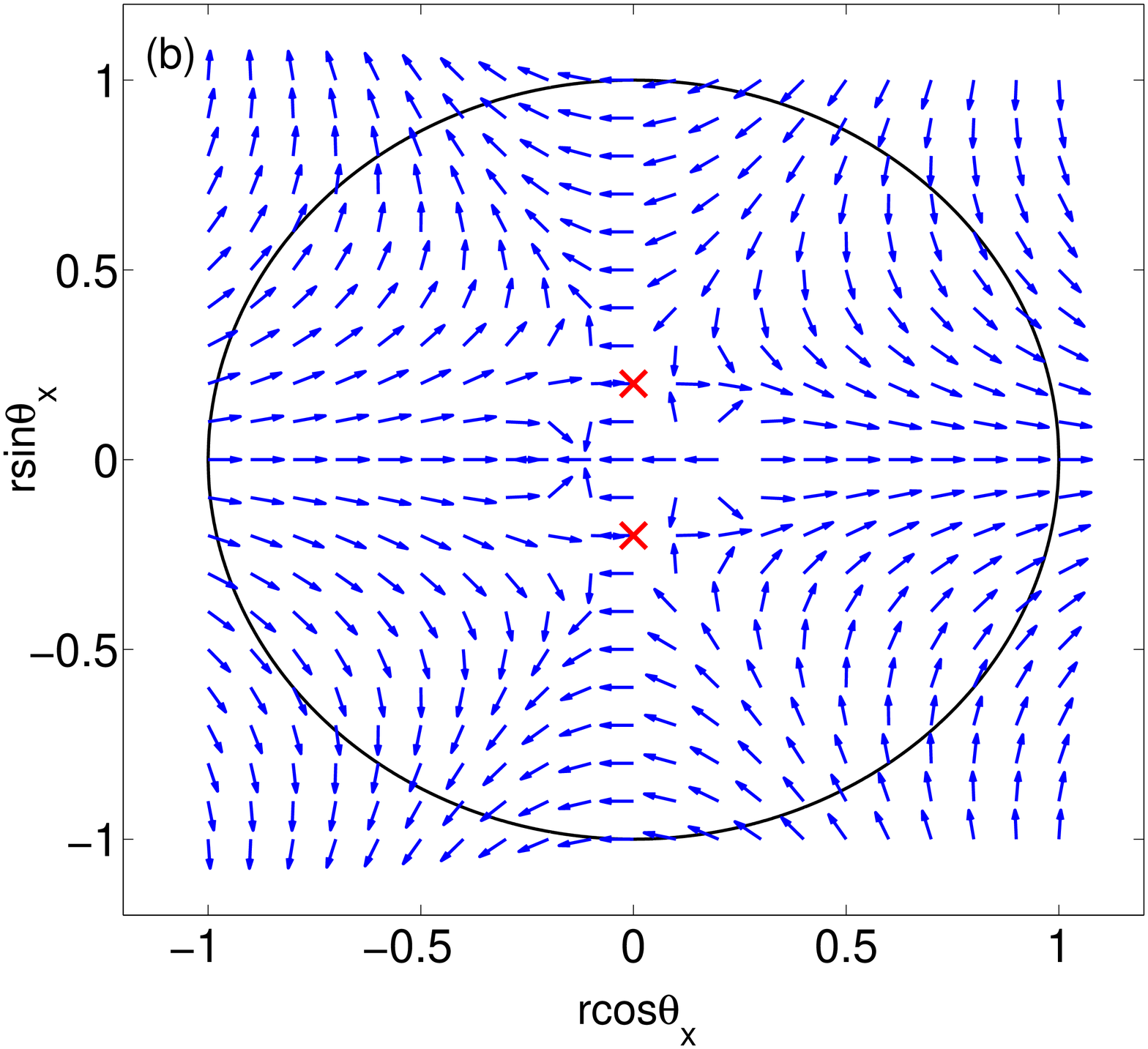}
\end{center}
\caption{Schematic picture of the $\phi_y(r,\theta_x)$
 configuration for the spin Hall system when (a) $c=0$;
 (b) $c\neq 0$. The direction of the arrows stands for
 the phase $\phi_y$. The crosses indicate the position
 of singularities and the circle is the unit circle $r=1$.}
\label{cherndist}
\end{figure}

If $r_1\neq r_2$, there are 4 gapless points $(0,r_{1,2}),
(\pi,r_{1,2})$ in the parameter space. In other words, the original
singularity at $r=0$ with winding number $4$ is split to $4$
singularities each with unit winding number. However, the time
reversal invariance will force $r_1=r_2$, since $\phi_{kL(R)1}$ are
the T-partner of $\phi_{-kL(R)2}$. Therefore, the 4 singularities
must merge to $2$, each with winding number 2, which is consistent
with the above numerical results. The schematic picture of
$\phi_y(r,\theta_x)$ configurations for $c=0$ and $c\neq 0$ is shown
in Fig. \ref{cherndist}.

Concluding this section, we would like to emphasize that the
existence of gapless edge states for a proper set of twisted
boundary conditions with inter-edge tunnelling should not be viewed
as a trivial result of fine tuning. For example, if one replaces the
inter-edge tunnelling term in Eq. (\ref{perturbHamiltonian}) by
\[
r\sum_s\int dk\left(\phi_{kLs\alpha}^\dagger
e^{i\theta}\phi_{kRs\alpha}+h.c.\right) \ ,
\]
the Hamiltonian will be gapped for any $(r,\theta)$ as long as
$u_{L,Rs}\neq 0$ and ${\rm Im}\left(u_{Ls}/u_{Rs}\right)\neq 0$, a
direct consequence of the vanishing Chern number in the charge
channel.

\section{Role of Time Reversal Invariance}

Time reversal invariance plays an important role in the topological
description of the quantum spin Hall systems. For the
non-interacting fermion case, a $Z_2$ classification has been
proposed in Ref.\cite{kane2005b}, namely the gaplessness of edge
states in open-boundary systems is protected by Kramers degeneracy,
if there are in total an odd number pairs of edge states. However,
the situation is very subtle when interactions are included in the
many-body system. Below we will first show a consequences of time
reversal invariance (Theorem 2) in a general many-body system in
subsection A, and then discuss about the $Z_2$-classification in
subsection B, where new predictions will be made based on Theorem 1.

\subsection{Conditions for Winding Number Doubling}

For simplicity, we consider the case when $\Gamma_x,\Gamma_y$ each
carries a simple representation of time reversal transformation:
\begin{eqnarray}
T\Gamma_\mu T^{-1}=\eta_\mu \Gamma_\mu,\qquad \eta_\mu=\pm 1 \ .
\end{eqnarray}

If the original Hamiltonian (\ref{originalH}) is TR-invariant, then
the TR property of the twisted Hamiltonian
(\ref{twistedHamiltonian}) will be
\begin{eqnarray}
TH_{\Gamma_x\Gamma_y}(r,\theta_x,\theta_y)T^{-1}
=H_{\Gamma_x\Gamma_y}(r,-\eta_x\theta_x,-\eta_y\theta_y)
\end{eqnarray}
Thus the gauge field has the symmetry property
\begin{eqnarray}
&&A_\mu(r,\theta_x,\theta_y)=-i\left\langle
G(r,\theta_x,\theta_y)\right|\partial_\mu\left|
G(r,\theta_x,\theta_y)\right\rangle\nonumber\\
&&=i\left\langle G(r,-\eta_x\theta_x,-\eta_y\theta_y)\right|
\partial_\mu\left|G(r,-\eta_x\theta_x,-\eta_y\theta_y)
\right\rangle\nonumber\\
&&=\eta_\mu
A_\mu(r,-\eta_x\theta_x,-\eta_y\theta_y) ,\nonumber\\
\Rightarrow && F_{xy}(r,\theta_x,\theta_y)= -\eta_x\eta_y
F_{xy}(r,-\eta_x\theta_x,-\eta_y\theta_y)\ .
\end{eqnarray}
Thus the Chern number $N$ is non-vanishing only when
$\eta_x\eta_y=-1$. In the usual QHE case $\eta_x=\eta_y=1$, which
means that the Hall conductivity vanishes for TR-invariant systems.
On the other hand, the spin Hall effect corresponds to $\eta_x=-1$,
$\eta_y=1$, since $\Gamma_x$ is a spin operator and $\Gamma_y=1$ is
taken to be the charge operator. Below we will focus on the case
with $\eta_x=-1,\eta_y=1$.

In general the first Chern number $N$ can be either even or odd
integer. However, we will show that $N$ must be even when
\begin{eqnarray}
e^{i\Gamma_x\pi}=-1 \ . \label{oddspin}
\end{eqnarray}
When $\Gamma_x$ is a component of the spin operator, this condition
simply means that the spin is a half-odd integer.

To prove this statement, it is convenient to introduce a deformed
Hamiltonian by redefining the twisted boundary condition in the
$x$-direction as
\begin{eqnarray}
t_{ij}\rightarrow t_{ij}re^{i\Gamma_x\theta_x}e^{i\alpha_x}, \qquad
\left\langle ij\right\rangle \text{ across }L_x \, \label{twist3}
\end{eqnarray}
and leaving the twisted boundary condition in the $y$-direction
unchanged as in Eq.(\ref{twist}). In this way, we denote the
auxiliary Hamiltonian as
\begin{eqnarray}
H'_{\Gamma_x\Gamma_y}\left(r,\theta_x,\alpha_x,\theta_y\right).
\end{eqnarray}
The introduction of the charge twist phase $e^{i\alpha_x}$ changes
the TR property of $H'$ to
\begin{eqnarray}
TH'_{\Gamma_x\Gamma_y}\left(r,\theta_x,\alpha_x,\theta_y\right)T^{-1}
\nonumber\\
=H'_{\Gamma_x\Gamma_y}\left(r,\theta_x,-\alpha_x,-\theta_y\right)\ .
\label{TRofHprim}
\end{eqnarray}
On the other hand, noticing the condition (\ref{oddspin}), we get
\begin{eqnarray}
&& H'_{\Gamma_x\Gamma_y}\left(r,\theta_x,\alpha_x,\theta_y\right)
\nonumber \\
&=& H'_{\Gamma_x\Gamma_y}\left(r,\theta_x+\pi,\alpha_x+\pi,
\theta_y\right). \label{oddspinHprim}
\end{eqnarray}
Combining Eqs. (\ref{TRofHprim}) and (\ref{oddspinHprim}) and taking
$\alpha_x=\pi/2$, we obtain
\begin{eqnarray}
TH'_{\Gamma_x\Gamma_y}\left(r,\theta_x,\frac\pi
2,\theta_y\right)T^{-1}
\nonumber \\
=H'_{\Gamma_x\Gamma_y}\left(r,\theta_x+\pi,\frac\pi
2,-\theta_y\right) \ . \label{oddspinSymmetry}
\end{eqnarray}

In the same way as for the Hamiltonian (\ref{twistedHamiltonian}),
the Berry phase gauge field $A_\mu'$ and the phase factor
$\phi'(r,\theta_x)=\int_0^{2\pi}A_y'(r,\theta_x,\theta_y)d\theta_y$
can be defined for (the unique ground state of) $H'$. Then Eq.
(\ref{oddspinSymmetry}) leads to the symmetry property
\begin{eqnarray}
\phi'(r,\theta_x)=\phi'(r,\theta_x+\pi), \label{centersymmetry}
\end{eqnarray}
which means $\phi'$ is symmetric under the central reflection
$(r,\theta)\rightarrow (r,\theta+\pi)$. Consequently, the positions
of the singularities are also symmetric under the same
transformation, thus the winding number must be even:
\begin{eqnarray}
N'(r)&\equiv&\int_0^{2\pi}\frac{\partial\phi'(r,\theta_x)}
{\partial \theta_x}d\theta_x \nonumber\\
&=&2\int_0^\pi\frac{\partial
\phi'(r,\theta_x)}{\partial\theta_x}d\theta_x=\text{even integer}\,
\label{evenchernnumber}
\end{eqnarray}
if the Hamiltonian
$H'_{\Gamma_x\Gamma_y}(r,\theta_x,\frac{\pi}2,\theta_y)$ is gapped
for all $\theta_x,\theta_y\in[0,2\pi)$ at given $r$.

As the last step, the no-phase-stiffness condition is introduced
again, which means the energy spectrum of the system is insensitive
to $\alpha_x$ in the thermodynamic limit. Therefore, whenever
$H'_{\Gamma_x\Gamma_y}(r,\theta_x,\frac{\pi}2,\theta_y)$ is gapped,
the Hamiltonian (\ref{twistedHamiltonian}),
\[
H_{\Gamma_x\Gamma_y}(r,\theta_x,\theta_y)
=H'_{\Gamma_x\Gamma_y}(r,\theta_x,0,\theta_y),
\]
should also be gapped. This implies that the singularity
distribution of $\phi(r,\theta_x)$ is the same as that of
$\phi'(r,\theta_x)$. As a result, the Chern number of the spin Hall
system $H_{\Gamma_x\Gamma_y}(r,\theta_x,\theta_y)$ must be an even
integer, and we have finally proved the following theorem:
\begin{itemize}
\item{\underline{Theorem 2}: If the conditions below are
satisfied:
\begin{enumerate}
\item{ $T\Gamma_xT^{-1}=-\Gamma_x,T\Gamma_yT^{-1}=\Gamma_y$,}
\item{ $\exp(i\Gamma_x\pi)=-1$,} \item{the original Hamiltonian is
TR-invariant, with charge conservation but without phase stiffness
(i.e. without superconductivity),}
\end{enumerate}
then the singularities in the phase field $\phi(r,\theta_x)$ defined
in Eq. (\ref{defineBerryphase}) distribute symmetrically with
respect to the origin $r=0$ on the $(r,\theta_x)$-plane, and the
Chern number for the unique ground state of
$H_{\Gamma_x\Gamma_y}(r,\theta_x,\theta_y)$ is an even integer. }
\end{itemize}

\subsection{Comments on $Z_2$ Classification}

After studying the general consequence of time reversal invariance,
now it is time to consider whether there is any additional
topological classification besides the U(1) Chern number for
TR-invariant systems. Although the singularities distribute
symmetrically around the origin $r=0$ when the conditions in Theorem
2 are satisfied, it is still possible for all singularities to move
away from the origin $r=0$, which means the edge states in the
open-boundary system is not protected to be gapless. However, it has
been shown analytically\cite{kane2005a,kane2005b} and
numerically\cite{sheng2005} that gapless edge states along the open
boundaries in a spin Hall insulator model proposed for graphene
remains robust against disorder. In Ref.\cite{sheng2006} it is
further shown that the existence of gapless edge states and the
non-vanishing bulk Chern number $N=2$ in the spin channel are in
one-to-one correspondence. The robustness of the gapless edge states
in an open-boundary system comes from Kramers degeneracy, and
applies only when there are an odd number of Kramers pairs on each
edge, which corresponds to the Chern number $N=4n+2,n\in
\mathbb{Z}$. In this way, a $Z_2$ classification is well-defined for
non-interacting fermion systems, which distinguishes the spin Hall
insulators with Chern number $N=4n+2$ from those with $N=4n$.

To generalize this mechanism to an interacting system, it should be
noticed that the condition for Kramers degeneracy,
\begin{eqnarray}
TTc_{k\alpha}T^{-1}T^{-1}=-c_{k\alpha} \label{Todd},
\end{eqnarray}
is a single-particle property, not a condition on the many-body wave
function. When interactions are taken into account, the gapless edge
states emerging at some singular point $(r,\theta_x)$ in parameter
space may carry quantum numbers different from the constituent
particle. Thus, the protection for Kramers degeneracy can survive
only if the condition (\ref{Todd}) is still true for the edge modes.
Such a protection can be understood both by edge dynamics and by
bulk topology. Below we will discuss about $Z_2$ classification from
the bulk topology as a consequence of both Kramers degeneracy and
Theorem 2.

To proceed, we introduce a conjecture that is known to be true for
non-interacting systems\cite{Hatsugai1993B}, but has not been proven
for a general many-body system:
\begin{itemize}
\item{\underline{Conjecture}: If there is a singularity of the
$\phi(r,\theta_x)$ field with winding number $N$ at some point
$(r,\theta_x)$, the corresponding Hamiltonian
$H_{\Gamma_x\Gamma_y}(r,\theta_x,\theta_y)$ has $2N$ branches of
gapless edge states. }
\end{itemize}

Below we will take this conjecture as the starting point and leave
its proof to future work. In accordance with this conjecture, a
singularity with winding number $N=1$ corresponds to $2$ branches of
edge modes. If the edge modes carry $T^2=-1$, the two branches must
be $T$-conjugate to each other. Therefore there must be a Kramers
degeneracy between them that cannot be broken by any perturbation
without breaking the time reversal symmetry. Recall that the twist
phase factor $e^{i\Gamma_x\theta_x}$ is T-even when
$T\Gamma_xT^{-1}=-\Gamma_x$, the local variation of parameter
$(r,\theta_x)$ around the singular point can be considered as local
perturbation respecting T-symmetry, which thus must maintain the
system gapless. In other words, the system cannot have point-like
singularity with winding number $N=1$ {\em anywhere in the parameter
space}. A similar argument shows that all the point-like singularity
in the ($r,\theta_x$)-plane with $N$ odd is forbidden. Consequently,
all the vortex-like singularities of the $\phi(r,\theta_x)$ field
has an even winding number. Taking into account the central
reflection symmetry of the singular points as a corollary of Theorem
2, we know that all the singularities away from $r=0$ provide a
winding number $4n$. Thus at least $2$ pairs of gapless edge states
must exist for an open-boundary system when the bulk Chern number is
$N=4n+2,\ (n\in\mathbb{Z})$.

In summary, the $Z_2$ classification is applicable when the system
satisfies all the conditions of Theorem 2 and also has {\em all its
low-energy quasiparticles odd under $T^2$-operation}. It should be
clarified that for an interacting system, the condition
(\ref{oddspin}) {\em does not} imply $T^2=-1$ for low-lying modes.
Since the properties of low-energy quasiparticles may change without
closing the bulk gap, there seems no reason to believe the $Z_2$
``order" is protected by the bulk topology alone; it rather depends
on the detail of interactions.

From the discussion above, new predictions can be obtained for the
$N=1$ system. In Ref.\cite{wu2005,xu2005}, the edge dynamics for
some $N=1$ systems has been studied, which shows that the edge
states in the open-boundary system can be gapped with proper
interactions. After such a gap is opened, there are two
possibilities for this system:
\begin{enumerate}
\item As shown in Ref.\cite{wu2005}, the ground state of an
open-boundary system may break time reversal symmetry spontaneously,
which leads to ground state degeneracy. In this case, the point
$r=0$ is still singular. If the Chern number $N=2$ is still carried
by this singularity, then no
other singularity necessarily exists.\\
\item If the singularity at $r=0$ does not carry any Chern number,
or if the TR-symmetry is recovered, then the bulk topology requires
the gapless edge states to exist with proper inter-edge tunnelling
strength. What's more, from the discussion in this subsection we
know that as long as the gapless edge states shift to $r\neq 0$,
they must be even under $T^2$ operation. Physically, this prediction
means that before the gap-opening transition, the edge modes are
$T^2$-odd quasi-particles, which carry similar quantum numbers as
free fermions; immediately after this transition, the edge modes
will become $T^2$-even, which may correspond to some kind of spin
and/or charge density waves.
\end{enumerate}

\section{Summary and Discussions}

In conclusion, in this paper we have established, for 2d insulators,
a general connection between bulk topological order and edge
dynamics illustrated by Theorem 1. The key idea behind demonstrating
this bulk-edge relation is to bring in an additional parameter that
describes the strength of edge tunnelling besides the twisted
boundary phases introduced in Ref.\cite{niu1985}. This results in a
three dimensional parameter space and implements a continuous
interpolation between a cylindrical system with open boundaries and
a toroidal system. In this framework, it is clarified that
generically the bulk topological order for an insulator in torus
geometry is not necessarily associated with gapless edge states in
cylinder geometry. On the one hand, the the emergence of gapless
edge states known for QH systems is shown to be closely related to
charge conservation in addition to topological reasons. On the other
hand, a QSH system with non-trivial bulk Chern number and gapped
edge states is predicted to have gapless edge states when inter-edge
tunnelling with proper strength is turned on. Finally, a general
consequence of time reversal invariance is deduced in Theorem 2. The
$Z_2$ classification is shown to be dependent, besides bulk
topology, also on the $T^2$ ($T$ being time reversal) quantum number
of relevant low-lying modes, which is not topological in nature and
can be changed by interactions without changing the bulk topological
order.

The proof of Theorem 1 can in principle be generalized to
topological insulators in higher dimensions, for which the
topological order is related to non-Abelian Berry phase gauge field
and higher Chern numbers. For example, in the four dimensional
quantum Hall effect\cite{zhang2001}, the second Chern number can be
defined when the Berry phase gauge field is ${\rm SU(N)}$ and the
parameter space is 4-dimensional as labeled by
$(\theta_x,\theta_y,\theta_z,\theta_w)$. In the same way as in Sec.
II, an amplitude $r$ can be introduced for any boundary and a
non-Abelian Berry phase field $U(r,\theta_\mu)\in {\rm SU(N)}$ can
be defined by a straightforward generalization of Eq.
(\ref{defineBerryphase}) to a Wilson loop operator:
\begin{eqnarray}
\Phi(r,\theta_\mu)=P\exp\left[i\int
A_w(r,\theta_x,\theta_y,\theta_z,\theta_w)d\theta_w\right],
\end{eqnarray}
in which $\mu=x,y,z$ and $P$ stands for path ordering. The second
Chern number is then identified as the winding number of the $U$
field on the 3-torus $T^3$ at $r=1$. By the same topological
reasoning, there must be singularities of the $U$ field for some
$(r,\theta_\mu)$ with $r<1$ when the bulk Chern number for the
system with $r=1$ is non-trivial, which corresponds to gapless edge
states in higher dimensional sense. In this way, Theorem 1 provides
a universal {\em modus operandi} to understand the bulk-edge
relationship in topological insulators.

This work is supported in part by the NSFC through the grants No.
10374058, and by the US NSF through the grants PHY-0457018 and
DMR-0342832, and by the US Department of Energy, Office of Basic
Energy Sciences under contract DE-AC03-76SF00515. We would like to
acknowledge helpful discussions with B. A. Bernevig, J. P. Hu, C.
Kane, J. Moore, D. N. Sheng, Z. Y. Weng, C. J. Wu and C. K. Xu.

\bibliography{U1}
\end{document}